\documentclass[12pt,a4paper,final]{iopart}
\usepackage{iopams}  
\usepackage{graphicx}
\usepackage[breaklinks=true,colorlinks=true,linkcolor=blue,urlcolor=blue,citecolor=blue]{hyperref}

\usepackage{graphicx}
\usepackage{dcolumn}
\usepackage{bm}
\usepackage{xfrac}
\usepackage{tikz}
\usetikzlibrary{decorations.pathreplacing, shadows, shapes}
\usepackage{multirow} 
\usepackage{setspace}
\makeatletter
\newcommand{\seq}{%
  \settowidth{\@tempdima}{-}
 \; \resizebox{\@tempdima}{\height}{=} \;%
}
\makeatother
\usepackage{subfigure}
\definecolor{mydarkred}{RGB}{153, 0,0} 
\definecolor{mydarkblue}{RGB}{0,128,128}
\definecolor{mygreen}{RGB}{175,233,198}
\definecolor{mydarkgreen}{RGB}{5,89,3}
\definecolor{mydarkgreen2}{RGB}{8,74,45}
\definecolor{myblue}{RGB}{175,221,233}
\definecolor{mylightblue}{RGB}{169,199,252}
\definecolor{myblue2}{RGB}{190,190,240}
\definecolor{myblue3}{RGB}{150,210,250}
\definecolor{mydarkblue}{RGB}{0,128,128}
\definecolor{mypurple}{RGB}{239,229,247}
\definecolor{mydarkpurple}{RGB}{179,0,250}
\definecolor{myyellow}{RGB}{255,238,170}
\definecolor{mylightbrown}{RGB}{209,166,102}
\definecolor{mydarkbrown}{RGB}{102,51,0}
\definecolor{mydarkred}{RGB}{153, 0,0} 
\definecolor{mydarkorange}{RGB}{229,114,0}
\definecolor{mylightorange}{RGB}{252,248,236}
\definecolor{mygray}{RGB}{240,238,228}
\definecolor{mydarkgray}{RGB}{50, 50, 50}
\definecolor{mygraywhite}{RGB}{243, 243,243}
\definecolor{mylightred}{RGB}{255, 147,147}
\definecolor{mylightpink}{RGB}{255, 51,153}
\definecolor{mylightorange}{RGB}{255, 189,91}
\newcommand{\ket}[1]{| #1 \rangle} 
\newcommand{\bra}[1]{\langle #1 |} 

\newcommand{\x}{\mathbf{x}}
\definecolor{mydarkblue}{RGB}{0,128,128}
\definecolor{mydarkred}{RGB}{153, 0,0} 

\begin{document}

\title{Quantum ensembles of quantum classifiers}

\author{Maria Schuld}
\address{Quantum Research Group, School of Chemistry and Physics, University of KwaZulu-Natal, Durban 4000, South Africa}
\ead{schuld@ukzn.ac.za}

\author{Francesco Petruccione$^{1,2}$}
\address{$^1$Quantum Research Group, School of Chemistry and Physics, University of KwaZulu-Natal, Durban 4000, South Africa}
\address{$^2$National Institute for Theoretical Physics}
\ead{petruccione@ukzn.ac.za}

\begin{abstract}
Quantum machine learning witnesses an increasing amount of quantum algorithms for data-driven decision making, a problem with potential applications ranging from automated image recognition to medical diagnosis. Many of those algorithms are implementations of quantum classifiers, or models for the classification of data inputs with a quantum computer. Following the success of collective decision making with ensembles in classical machine learning, this paper introduces the concept of quantum ensembles of quantum classifiers. Creating the ensemble corresponds to a state preparation routine, after which the quantum classifiers are evaluated in parallel and their combined decision is accessed by a single-qubit measurement. This framework naturally allows for exponentially large ensembles in which -- similar to Bayesian learning -- the individual classifiers do not have to be trained. As an example, we analyse an exponentially large quantum ensemble in which each classifier is weighed according to its performance in classifying the training data, leading to new results for quantum as well as classical machine learning.
\end{abstract}

\vspace{2pc}

\section{Introduction}
In machine learning, a classifier can be understood as a mathematical model or computer algorithm that takes input vectors of features and assigns them to classes or `labels'. For example, the features could be derived from an image and the label assigns the image to the classes ``shows a cat'' or ``shows no cat''. Such a classifier can be written as a function $f: \mathcal{X} \rightarrow \mathcal{Y}$ mapping from an input space $\mathcal{X}$ to the space of possible labels $\mathcal{Y}$. The function can depend on a set of parameters $\theta$, and training the classifier refers to fitting the parameters to sample data of input-label pairs in order to get a model $f(\x; \theta), \x \in \mathcal{X}$ that generalises from the data how to classify future inputs. It is by now common practise to consult not only one trained model but to train an ensemble of models $f(\x; \theta_i)$, $i = 1,...,E$ and derive a collective prediction that supersedes the predictive power of a single classifier \cite{dietterich00}.\\

In the emerging discipline of quantum machine learning, a number of quantum algorithms for classification have been proposed \cite{rebentrost14, lloyd13, amin16} and demonstrate how to train and use models for classification on a quantum computer. It is an open question how to cast such quantum classifiers into an ensemble framework that likewise harvests the strengths of quantum computing, and this article is a first step to answering this question. We will focus on a special type of quantum classifier here, namely a model where a set of parameters $\theta$ can be expressed by a $n$-qubit state $\ket{\theta}$, and the classification result is encoded in a separate register of qubits. This format allows us to create a `superposition of quantum models' $\sum_{\theta}\ket{\theta}$ and evaluate them in parallel. A state preparation scheme can be used to weigh each classifier, thereby creating the ensemble. This allows for the instantaneous evaluation of exponentially large quantum ensembles of quantum classifiers.\\

Exponentially large ensembles do not only have the potential to increase the predictive power of single quantum classifiers, they also offer an interesting perspective on how to circumvent the training problem in quantum machine learning. Training in the quantum regime relies on methods that range from sampling from quantum states \cite{amin16} to quantum matrix inversion \cite{rebentrost14} and Grover search \cite{kapoor16}. However, for complex optimisation problems where little mathematical structure is given (an important example being feed-forward neural networks), the translation of iterative methods such as backpropagation to efficient quantum algorithms is less straight forward (see Ref. \cite{rebentrost16grad}). It is known from classical machine learning that one can avoid optimisation by exchanging it for integration: In \textit{Bayesian learning} one has to solve an integral over all possible parameters instead of searching for a single optimal candidate. The idea of integrating over parameters can be understood as forming a collective decision by consulting all possible models from a certain family and weigh them according to their desired influence -- which is the approach of the ensemble framework. In other words, quantum ensembles of quantum classifiers offer an interesting perspective to optimisation-free learning with quantum computers.\\

In order to illustrate the concept of quantum ensembles of quantum classifiers, we investigate an exponentially large ensemble inspired by Bayesian learning, in which every ensemble member is weighed according to its accuracy on the data set. We give a quantum circuit to prepare such a quantum ensemble for general quantum classifiers, from which the collective decision can be computed in parallel and evaluated from a single qubit measurement. It turns out that for certain models this procedure effectively constructs an ensemble of accurate classifiers (those that perform on the training set better than random guessing). It has been shown that in some cases, accurate but weak classifiers can build a strong classifier, and analytical and numerical investigations show that this may work. To our knowledge, this result has not been established in the classical machine learning literature and shows how quantum machine learning can stimulate new approaches for traditional machine learning as well. \\

This article is structured as follows. Section \ref{Sec2} provides definitions and summarises research relevant to our work, while Section \ref{Sec3} introuduces the formalism of quantum ensembles of quantum classifiers. Section \ref{Sec4} presents the example of an accuracy-weighted ensemble, and \ref{Sec5} shows that for some cases this is an effective ensemble of accurate members. Section \ref{Sec6} analyses the example analytically and leads to the conclusion in Section \ref{Sec7}. 

\section{Classification with asymptotically large ensembles of accurate models} \label{Sec2}

Before introducing the quantum ensemble framework as well as the example of an accuracy-weighed ensemble, we provide a review of classical results on ensemble methods. An emphasis lies on studies that analyse asymptotically large ensembles of accurate classifiers, and although no rigorous proofs are available, we find strong arguments for the high predictive power of large collections of weak learners. The problem we will focus on here is a supervised binary pattern classification task. Given a dataset $\mathcal{D}=\{(\mathbf{x}^{(1)}, y^{(1)}),...,(\mathbf{x}^{(M)},y^{(M)})\}$ with inputs $\mathbf{x}^{(m)} \in \mathbb{R}^N$ and outputs or labels $y^{(m)} \in \{-1,1\}$ for $ m=1,...,M$, as well as a new input $\tilde{\mathbf{x}}$. The goal is to predict the unknown label $\tilde{y}$. Consider a classifier 
\begin{equation} y = f(\mathbf{x}; \theta), \label{Eq:model}\end{equation}
with input $\mathbf{x} \in \mathcal{X}$ and parameters $\theta$. As mentioned above, the common approach in machine learning is to choose a model by fitting the parameters to the data $\mathcal{D}$. Ensemble methods are based on the notion that allowing only one final model $\theta$ for prediction, whatever intricate the training procedure is, will neglect the strengths of other candidates even if they have an overall worse performance. For example, one model might have learned how to deal with outliers very well, but at the expense of being slightly worse in predicting the rest of the inputs. This `expert knowledge' is lost if only one winner is selected. The idea is to allow for an ensemble or committee $\mathcal{E}$ of trained models (sometimes called `experts' or `members') that take the decision for a new prediction together. Considering how familiar this principle is in our societies, it is surprising that this thought only gained widespread attention as late as the 1990s. \\

\begin{figure}[t]
\centering
\begin{tikzpicture}
\path (3,0) node[anchor=center, align = center, rectangle, fill = white](sp) {\textit{selection}\\\textit{ procedure}};
 \draw [- ](4.3,2)--(4,2)-- (4,-2)--(4.3,-2); 
\path (5,1.5) node[anchor=center, align = center, rectangle, fill = mypurple] (m1){$f(x;\theta_{1})$};
\path (5,0.5) node[anchor=center, align = center, rectangle, fill = mypurple](m2) {$f(x;\theta_{2})$};
\path (5,-0.5) node[anchor=center, align = left, rectangle, fill = white] (md){$\vdots$};
\path (5,-1.5) node[anchor=center, align = center, rectangle, fill = mypurple](mN) {$f(x;\theta_{N})$};
 \draw [-> ](m1)-- (8,0); 
 \draw [-> ](m2)-- (8,0); 
 \draw [-> ](mN)-- (8,0); 
\path (7,-1.5) node[anchor=center, align = center, rectangle,circle, draw, fill = mylightbrown] (x){$\tilde{x}$};
\path (7.8,0) node[anchor=center, align = center, rectangle, fill = white] (dp){\textit{ decision}\\\textit{  procedure}};
\path (9.6,0) node[anchor=center, align = center, rectangle,circle, draw, fill = myyellow] (y){$\tilde{y}$};
 \draw [-> ](dp)-- (y); 
 \draw [-> ](x)-- (dp);
\path (7.8,0) node[anchor=center, align = center, rectangle, fill = white] (dp){\textit{ decision}\\\textit{  procedure}}; 
\path (3,1.5) node[anchor=center, align = center, rectangle,circle, draw, fill = mygreen] (D){$\mathcal{D}$};
\draw[->] (D)--(sp);
\path (7.8,1.5) node[anchor=center, align = center, rectangle,circle, draw, fill = mygreen] (D){$\mathcal{D}$};
\draw[->] (D)--(dp);
\end{tikzpicture}
\caption{The principle of ensemble methods is to select a set of classifiers and combine their predictions to obtain a better performance in generalising from the data. Here, the $N$ classifiers are considered to be parametrised functions from a family $\{f(x ; \theta)\}$, where the set of parameters $\theta$ solely defines the individual model. The dataset $\mathcal{D}$ is consulted in the selection procedure and sometimes also plays a role in the decision procedure where a label $\tilde{y}$ for a new input $\tilde{x}$ is chosen.}
\label{Fig:idea}
\end{figure}
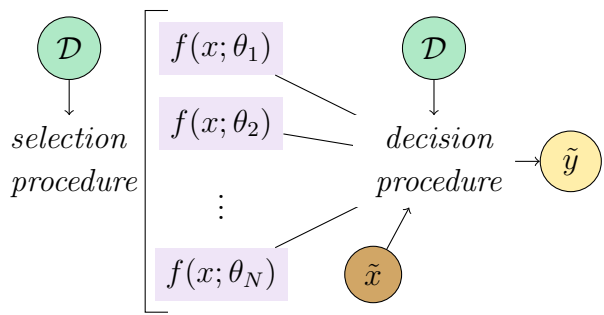

Many different proposals have been put forward of how to use more than one model for prediction. The proposals can be categorised along two dimensions \cite{kuncheva04}, first the selection procedure they apply to obtain the ensemble members, and second the decision procedure defined to compute the final output (see Figure \ref{Fig:idea}). Note that here we will not discuss ensembles built from different machine learning methods but consider a parametrised model family as given by Eq. (\ref{Eq:model}) with fixed hyperparameters. An example is a neural network with a fixed architecture and adjustable weight parameters. A very straight-forward strategy of constructing an ensemble is to train several models and decide according to their majority rule. More intricate variations are popular in practice and have interesting theoretical foundations. \textit{Bagging} \cite{breiman01} trains classifiers on different subsamples of the training set, thereby reducing the variance of the prediction. \textit{AdaBoost} \cite{schapire90, freund95} trains subsequent models on the part of the training set that was misclassified previously and applies a given weighing scheme, which can be understood to reduce the bias of the prediction. \textit{Mixtures of experts} \cite{jacobs91} train a number of classifiers using a specific error function and in a second step train a `gating network' that defines the weighing scheme. For all these methods, the ensemble classifier can be written as
\begin{equation} \tilde{y} = \mathrm{sgn} \left( \sum\limits_{\theta \in \mathcal{E}} w_{\theta} f(\tilde{x};\theta) \right).\label{Eq:ens_ens}
\end{equation}
The coefficients $w_{\theta}$ weigh the decision $f(\tilde{x};\theta) \in \{-1,1\}$ of each model in the ensemble $\mathcal{E}$ specified by $\theta$, while the sign function assigns class $1$ to the new input if the weighed sum is positive and $-1$ otherwise. It is important for the following to rewrite this as a sum over all $E$ \textit{possible} parameters. Here we will use a finite number representation and limit the parameters to a certain interval to get the discrete sum
\begin{equation}\tilde{y} = \mathrm{sgn} \left( \sum\limits_{\theta = 0}^{E-1} w_{\theta} f(\tilde{x};\theta) \right). \label{Eq:ensclass}
\end{equation}
In the continuous limit, the sum has to be replaced by an integral. In order to obtain the ensemble classifier of Eq. (\ref{Eq:ens_ens}), the weights $w_{\theta}$ which correspond to models that are not part of the ensemble $\mathcal{E}$ are set to zero. Given a model family $f$, an interval for the parameters as well as a precision to which they are represented, an ensemble is therefore fully defined by the set of weights $\{w_0...w_{E-1}\}$. \\

Writing a sum over all possible models provides a framework to think about asymptotically large ensembles which can be realised by quantum parallelism. Interesting enough, this formulation is also very close to another paradigm of classification, the Bayesian learning approach \cite{ghahramani15, duda12}. Given a training dataset $\mathcal{D}$ as before and understanding $\mathbf{x}$ as well as $y$ as random variables, the goal of classification in the Bayesian setting is to find the conditional probability distribution 
\begin{equation} p(\mathbf{x},y | \mathcal{D}) =  \int p(\mathbf{x},y |\theta ) p(\theta|\mathcal{D} ) d\theta, \label{Eq:ml_bayesint}\end{equation}
from which the prediction can be derived, for example by a Maximum A Posteriori estimate. The first part of the integrand, $p(\mathbf{x},y |\theta )$, is the probability of an input-label pair to be observed given the set of parameters for the chosen model. 
The correspondence of Eq.  (\ref{Eq:ensclass})  and Eq. (\ref{Eq:ml_bayesint}) becomes apparent if one associates $f(\x;\theta)$ with $p(\mathbf{x},y |\theta ) $ and interprets $w_{\theta}$ as  an estimator for the posterior $p(\theta| \mathcal{D})$ of $\theta$ being the true model given the observed data. If we also consider different model families specified by the hyperparameters, this method turns into \textit{Bayesian Model Averaging} which is sometimes included in the canon of ensemble method (although being based on a rather different theoretical foundation \cite{minka00}). \\

\begin{figure}[t]
\centering
\includegraphics[width=0.45\textwidth]{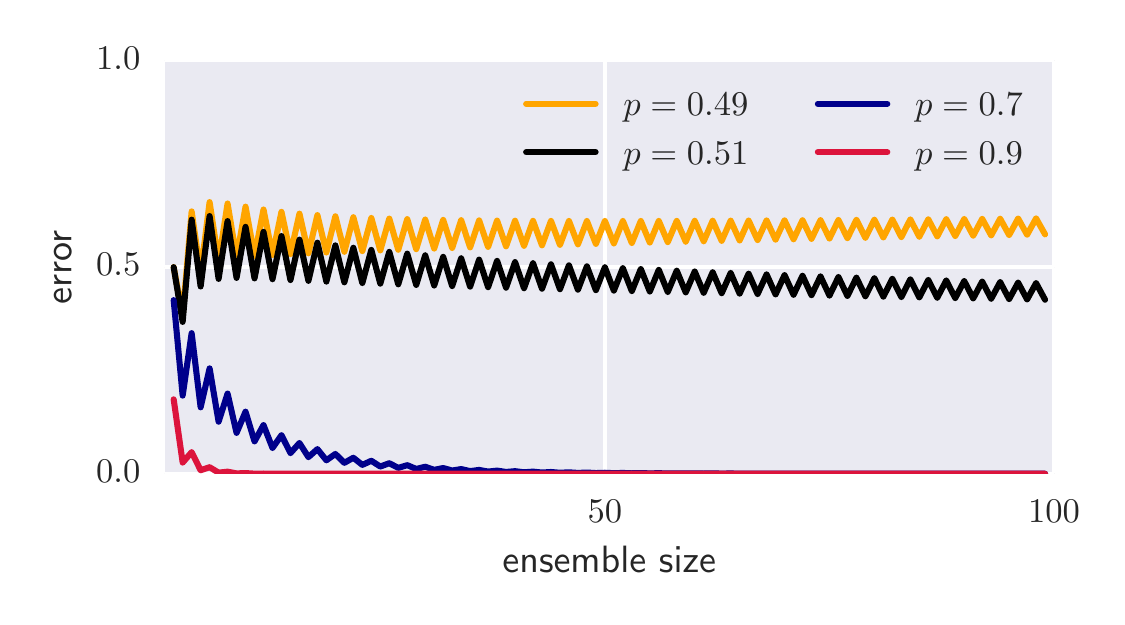}
\includegraphics[width=0.45\textwidth]{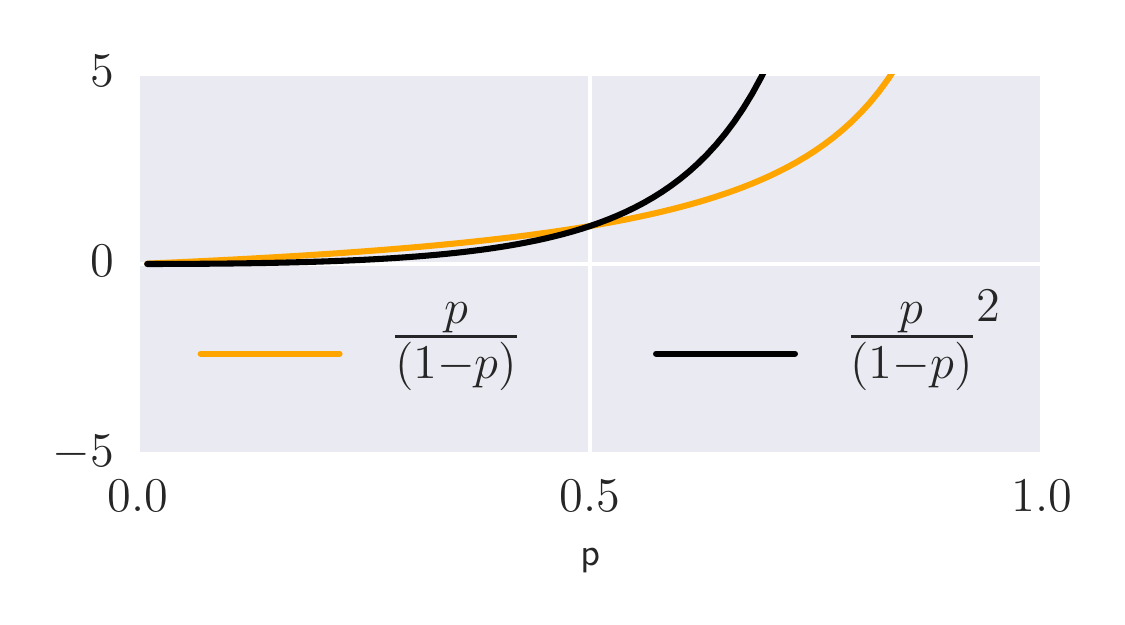}  
 \caption[Plot of the odds ratio and asymptotic prediction error]{Left: Prediction error when increasing the size of an ensemble of classifiers each of which has an accuracy $p$. Asymptotically, the error converges to zero if $p>0.5$.  Right:  For $p>0.5$, the odds ratio $p/(1-p)$ grows slower than its square. Together with the results from Lam et al described in the text, this is an indication that adding accurate classifiers to an ensemble has a high chance to increase its predictive power.}
\label{Fig:ens_oddsratio}
\end{figure}

Beyond the transition to a Bayesian framework, increasing the size of the ensemble to include all possible parameters has been studied in different contexts. In some cases  adding accurate classifiers has been shown to increase the performance of the ensemble decision. Accurate classifiers have an accuracy $a$ (estimated by the number of correctly classified test samples divided by the total samples of the validation set) of more than $0.5$, and are thus better than random guessing, which means that they have `learned' the pattern of the training set to at least a small extend. The most well-known case has been developed by Schapire \cite{schapire90} leading to the aforementioned AdaBoost algorithm where a collection of weak classifiers with accuracy slightly better than $0.5$ can be turned into a strong classifier that is expected to have a high predictive power. The advantage here is that weak classifiers are comparably easy to train and combine. But people thought about the power of weak learners long before AdaBoost. The \textit{Cordocet Jury Theorem} from 1758 states that considering a committee of judges where each judge has a probability $p$ with $p>0.5$ to reach a correct decision, the probability of a correct collective decision by majority vote will converge to $1$ as the number of judges approaches infinity. This idea has been applied to ensembles of neural networks by Hansen and Salamon \cite{hansen90}. If all ensemble members have a likelihood of $p$ to classify a new instance correctly, and their errors are uncorrelated, the probability that the majority rule classifies the new instance incorrectly is given by
\[ \sum\limits_{k>E/2}^E \begin{pmatrix} E \\ k\end{pmatrix} p^{E-k}(1-p)^{k}, \]
where $E$ is again the ensemble size. The convergence behaviour is plotted in Figure \ref{Fig:ens_oddsratio} (left) for different values of $p$. The assumption of uncorrelated errors is idealistic, since some data points will be more difficult to classify than others and therefore tend to be misclassified by a large proportion of the ensemble members.  Hansen and Salamon argue that for the highly overparametrised neural network models they consider as base classifiers, training will get stuck in different local minima, so that the ensemble members will be sufficiently diverse in their errors. \\

A more realistic setting would also assume that each model has a different prediction probability $p$ (that we can measure by the accuracy $a$), which has been investigated by Lam and Suen \cite{lam97}. The change in prediction power with the growth of the ensemble obviously depends on the predictive power of the new ensemble member, but its sign can be determined. Roughly stated, adding two classifiers with accuracies $a_1, a_2$ to an ensemble of size $2n$ will increase the prediction power if the value of $\frac{a_1a_2}{(1-a_1)(1-a_2)}$ is not less than the odds ratio $\frac{a_{i}}{(1-a_{i})}$ of any ensemble member, $i=1,...,E$. When plotting the odds ratio and its square in Figure \ref{Fig:ens_oddsratio} (right), it becomes apparent that for all $a_i>0.5$ chances are high to increase the predictive power of the ensemble by adding a new weak learner. Together, the results from the literature results suggest that constructing large ensembles of accurate classifiers can lead to a strong combined classifier. \\


Before proceeding to quantum models, another result is important to mention. If we consider all possible parameters $\theta$ in the sum of Equation (\ref{Eq:ensclass}) and assume that the model defined by $\theta$ has an accuracy $a_{\theta}$ on the training set, the optimal weighing scheme \cite{shapley84} is given by 
\begin{equation}
w(\theta) = \log \frac{a_{\theta}}{1-a_{\theta}}.
\label{Eq:quens_log}
\end{equation} 
It is interesting to note that this weighing scheme corresponds to the weights chosen in AdaBoost for each trained model, where they are derived from what seems to be a different theoretical objective.

\section{Quantum ensembles of quantum classifiers}\label{Sec3}
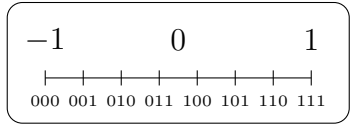
\begin{figure}
\centering
\begin{tikzpicture}
\draw[rounded corners=0.2cm] (10.5,1.4) rectangle (15,3);
\draw [](11,2)-- (14.5,2);
\draw [](11,2.1)-- (11,1.9);
\draw [](11.5,2.1)-- (11.5,1.9);\draw [](12,2.1)-- (12,1.9);\draw [](12.5,2.1)-- (12.5,1.9);\draw [](13,2.1)-- (13,1.9);\draw [](13.5,2.1)-- (13.5,1.9);\draw [](14,2.1)-- (14,1.9);\draw [](14.5,2.1)-- (14.5,1.9);
\path (11,2.5) node[anchor=center, align = center] { $-1$}; 
\path (14.5,2.5) node[anchor=center, align = center] { $1$}; 
\path (12.75,2.5) node[anchor=center, align = center] { $0$}; 
\path (11,1.7) node[anchor=center, align = center] { \tiny $000$}; 
\path (11.5,1.7) node[anchor=center, align = center] { \tiny $001$}; 
\path (12,1.7) node[anchor=center, align = center] { \tiny $010$}; 
\path (12.5,1.7) node[anchor=center, align = center] { \tiny $011$}; 
\path (13,1.7) node[anchor=center, align = center] { \tiny $100$}; 
\path (13.5,1.7) node[anchor=center, align = center] { \tiny $101$}; 
\path (14,1.7) node[anchor=center, align = center] {\tiny $110$}; 
\path (14.5,1.7) node[anchor=center, align = center] {\tiny $111$}; 
\end{tikzpicture}
\caption[Example for the representation of parameters in a uniform quantum superposition of three qubits.]{Example for the representation of parameters in a uniform quantum superposition of three qubits. Each computational basis state $\ket{000}, \ket{001},...,\ket{111}$ corresponds to a parameter in the interval $[-1,1]$. }
\label{Fig:ens_partition}
\end{figure}

The idea of this section is to cast the notion of ensembles into a quantum algorithmic framework, based on the idea that quantum parallelism can be used to evaluate ensemble members of an exponentially large ensemble in one step. Consider a quantum routine $\mathcal{A}$ which `computes' a model function $f(\x;\theta)$,
\[\mathcal{A} \; \ket{\x} \ket{\theta} \ket{0} \rightarrow \ket{\x} \ket{\theta} \ket{f(\x;\theta)}, \]
which we will call a \textit{quantum classifier} in the following. The last qubit $\ket{f(\x;\theta)}$ thereby encodes class $f(\x;\theta) = -1$ in state $\ket{0}$ and class $1$ in state $\ket{1}$. Note that it is not important whether the registers $\ket{\x}, \ket{\theta}$ encode the classical vectors $\x, \theta$ in the amplitudes or qubits of the quantum state. If encoding classical information into the binary sequence of computational basis states (i.e. $x =2 \rightarrow 010 \rightarrow \ket{010}$ ), every function $f(\x;\theta)$ a classical computer can compute efficiently could in principle be translated into a quantum circuit $\mathcal{A}$. This means that every classifier leads to an efficient quantum classifier (possibly with large polynomial overhead). An example for a quantum perceptron classifier can be found in Ref. \cite{schuld15perc}, while feed-forward neural networks have been considered in \cite{wan16}. With this definition of a quantum classifier, $\mathcal{A}$ can be implemented in parallel to a superposition of parameter states.
\[\mathcal{A} \; \ket{\x}  \;\frac{1}{\sqrt{E}}\sum_{\theta} \ket{\theta}\; \ket{0} \rightarrow \ket{\x} \; \frac{1}{\sqrt{E}}\sum_{\theta}  \ket{\theta} \; \ket{f(\x;\theta)}. \]
For example, given $\theta \in [a,b]$, the expression $\frac{1}{\sqrt{E}}\sum_{\theta} \ket{\theta}$ could be a uniform superposition, 
\[  \frac{1}{\sqrt{E}}\sum_{i =0}^{2^{\tau}-1} \ket{i},\]
where each computational basis state $\ket{i}$ corresponds to a $\tau$-bit length binary representations of the parameter, dividing the interval that limits the parameter values into $2^{\tau}$ candidates (see Figure \ref{Fig:ens_partition}).\\

As explained before, an ensemble method can be understood as a weighing rule for each model in the sum of Eq. (\ref{Eq:ensclass}). We will therefore require a second quantum routine, $\mathcal{W}$, which turns the uniform superposition into a non-uniform one,
\[\mathcal{W} \;\; \ket{\x} \; \frac{1}{\sqrt{E}}\sum_{\theta} \ket{\theta} \ket{0} \rightarrow \ket{\x}  \; \frac{1}{\sqrt{E \chi}}\sum_{\theta} \sqrt{w_{\theta}}  \ket{\theta} \ket{0}, \]
weighing each model $\theta$ by a classical probability $w_{\theta}$. We call this routine a \textit{quantum ensemble} since the weights define which model contributes to what extend to the combined decision. The normalisation factor $\chi$ ensures that $\sum_{\theta} \frac{w_{\theta}}{E\chi} =1$. Note that this routine can be understood as a state preparation protocol of a qsample.\\

Together, the quantum routines $\mathcal{A}$ and $\mathcal{W}$ define the quantum ensemble of quantum classifiers, by first weighing the superposition of models (in other words the basis states in the parameter register) and then computing the model predictions for the new input $\tilde{\x}$ in parallel. The final quantum state is given by 
\[  \ket{\tilde{\x}} \frac{1}{\sqrt{E\chi}}\sum_{\theta} \sqrt{w_{\theta}}  \ket{\theta} \ket{f(\tilde{\x};\theta)}.\]
The measurement statistics of the last qubit contain the ensemble prediction: The chance of measuring the qubit in state $0$ is the probability of the ensemble deciding for class $-1$ and is given by 
\[ p(\tilde{y} = -1) = \sum\limits_{\theta \in \mathcal{E}^+} \frac{w_{\theta}}{E\chi}, \]
while the chance of measuring the qubit in state $1$ reveals $p(\tilde{y}=1)$ and is  given by
\[p(\tilde{y} = 1) =\sum\limits_{\theta \in \mathcal{E}^-} \frac{w_{\theta}}{E\chi}, \]
where $\mathcal{E}^{\pm}$ is the subset of $\mathcal{E}$ containing only models with $f(\tilde{\x}; \theta) = \pm 1$. After describing the general template, we will now look at how to implement a specific weighing scheme with a quantum routine $\mathcal{W}$ for general models $\mathcal{A}$ and analyse the resulting classifier in more detail.

\section{Choosing the weights proportional to the accuracy}\label{Sec4}
As an illustrative case study, we choose weights that are proportional to the accuracy $a_{\theta}$ of each model, as measured on the training set. More precisely, this is the proportion of correct classifications over the number of data points
\[a_{\theta} =\frac{1}{M} \sum\limits_{m=1}^{M} \frac{1}{2} |f(x^{m}; \theta_i) + y^{m}|. \] 
Note that while usually an estimate for the accuracy of a trained model is measured on a separate test or validation set, we require the accuracy here to build the classifier in the first place and therefore have to measure it on the training set. The goal of the quantum algorithm is to prepare a quantum state where each model represented by state $\ket{\theta}$ has a probability to be measured that is proportional to its accuracy, $w_{\theta} \propto a_{\theta}$.\\

The weighing routine $\mathcal{W}$ can be constructed as follows. Required for computation is a system of $(\delta+1) + \tau + 1 + 1$ qubits divided into four registers: the data register, the parameter register, the output register and the accuracy register,
\begin{equation} \underbrace{\ket{0...0; 0}}_{\delta+1 \;\; \mathrm{qubits}}\; \otimes \; \underbrace{\ket{0...0}}_{\tau \;\; \mathrm{qubits}} \; \otimes \; \ket{0} \; \otimes \; \ket{0}. \label{Eq:quens1}\end{equation} 
Assume a quantum classification routine $\mathcal{A}$ is given. As a first step, $\tau$ Hadamards bring the parameter register into a uniform superposition, and a Hadamard is applied to the accuracy qubit:
\[  \frac{1}{\sqrt{E}}\sum\limits_{i=0}^{2^{\tau}-1} \ket{0...;0} \ket{i}\ket{0} \; \frac{1}{\sqrt{2}}(\ket{0} + \ket{1}). \]
Each $\ket{i}$ thereby encodes a set of parameters $\theta$. We now `load' the training pairs successively into the data register, compute the outputs in superposition by applying the core routine $\mathcal{A}$ and rotate the accuracy qubit by a small amount towards $\ket{0}$ or $\ket{1}$ depending on whether target output and actual output have the same value (i.e. by a XOR gate on the respective qubits together with a conditional rotation of the last qubit). The  core routine and loading step are then uncomputed by applying the inverse operations. \footnote{One could alternatively prepare a training superposition $\frac{1}{\sqrt{M}} \sum_m \ket{\x^m}$ and trace the training register out in the end. The costs remain linear in the number of training vectors times the bit-depth for each training vector for both strategies.}\\

After all training points have been processed, the accuracy qubit is entangled with the parameter register and in state $ \ket{q_{\theta}} = \sqrt{a_{\theta}} \ket{0} +\sqrt{1- a_{\theta}}\ket{1} $. A postselective measurement on the accuracy qubit only accepts when it is in in state $\ket{0}$ and repeats the routine otherwise. This selects the $\ket{0}$-branch of the superposition and leaves us with the state
\[  \frac{1}{\sqrt{E\chi}}\sum\limits_{\theta} \sqrt{a_{\theta}} \ket{0...;0} \ket{\theta} \ket{0} , \]
where the normalisation factor $\chi$ is equal to the acceptance probability  $p_{\mathrm{acc}} = \frac{1}{E }\sum_{\theta} a_{\theta} $. The probability of acceptance influences the runtime of the algorithm, since a measurement of  the ancilla in $1$ means we have to abandon the result and start the routine from scratch. We expect that choices of the parameter intervals and data pre-processing allows us to keep the acceptance probability sufficiently high for many machine learning applications, as most of the $a_{\theta}$ can be expected to be distributed around $0.5$. This hints towards rejection sampling as a promising tool to translate the accuracy-weighted quantum ensemble into a classical method. \\

Now load the new input into the first $\delta$ qubits of the data register, apply the routine $\mathcal{A}$  once more and uncompute (and disregard) the data register to obtain
\[ \ket{\psi} = \frac{1}{\sqrt{E\chi}}\sum\limits_{\theta} \sqrt{a_{\theta}} \ket{\theta} \ket{f(\tilde{\x},\theta)} . \]
The measurement statistics of the last qubit now contain the desired value. More precisely, the expectation value of $\mathbb{I} \otimes \sigma_z $ is given by
\[ \bra{\psi}\mathbb{I}\otimes \sigma_z\ket{\psi} =\frac{1}{E\chi} \sum\limits_{\theta} a_{\theta} (f(\tilde{x};\theta) + 1)/2 ,\]
and corresponds to the classifier in Eq. (\ref{Eq:ensclass}). Repeated measurements reveal this expectation value to the desired precision.\\

\section{Why accuracies may be good weights}\label{Sec5}

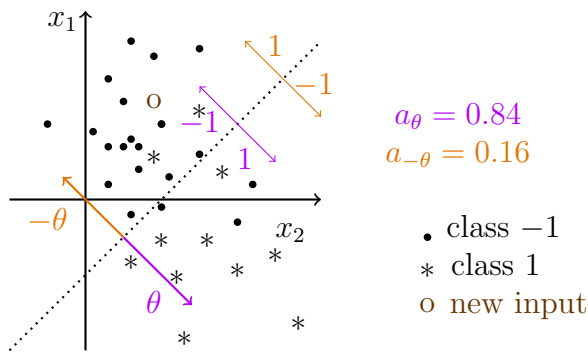
\begin{figure}[t]
\centering
\begin{tikzpicture}
 \draw [->, thick](-1,0)-- (3.1,0); 
 \draw [->,thick](0,-2)-- (0,2.5); 
\path (-0.3,2.3) node[anchor=base] {$x_1$};
\path (2.7,-0.5) node[anchor=base] {$x_2$};
\draw[dotted, thick] (-1,-2) -- (3.1,2.1);
\fill(1,1) circle (0.5mm);\fill(0.6,2.1) circle (0.5mm);\fill(1.5,2) circle (0.5mm);\fill(0.5,0.7) circle (0.5mm);\fill(-0.5,1) circle (0.5mm);\fill(0.3,1.6) circle (0.5mm);\fill(0.1,0.9) circle (0.5mm);\fill(0.9,1.9) circle (0.5mm);\fill(0.3,0.2) circle (0.5mm);\fill(0.5,1.3) circle (0.5mm);\fill(0.7,0.4) circle (0.5mm);\fill(0.6,-0.2) circle (0.5mm);\fill(2,-0.3) circle (0.5mm);\fill(2.2,0.2) circle (0.5mm);\fill(0.7,0.7) circle (0.5mm);\fill(1.5,0.6) circle (0.5mm);\fill(0.3,0.7) circle (0.5mm);\fill(0.6,0.8) circle (0.5mm);\fill(1.1,0.3) circle (0.5mm);\fill(1,1) circle (0.5mm);\fill(1,-0.1) circle (0.5mm);
\node[] at (2,-1) {*};\node[] at (1,-0.6) {*};\node[] at (1.8,0.3) {*};\node[] at (1.5,1.1) {*};\node[] at (2.5,-0.8) {*};\node[] at (2.8,-1.7) {*};\node[] at (1.3,-1.9) {*};\node[] at (1.6,-0.6) {*};\node[] at (1.2,-1.1) {*};\node[] at (0.6,-0.9) {*};\node[] at (0.9,0.5) {*};
\node[color = mydarkbrown] at (0.9,1.3) {o};
 \draw [->, mydarkorange](2.6,1.6)-- (3.1,1.1); 
 \draw [->, mydarkorange](2.6,1.6)-- (2.1,2.1); 
\path (3.0,1.4) node[anchor=base,mydarkorange] {$-1$};
\path (2.5,1.9) node[anchor=base,mydarkorange] {$1$};
 \draw [->, mydarkpurple ](2,1)-- (2.5, 0.5); 
 \draw [->, mydarkpurple ](2,1)-- (1.5,1.5); 
\path (2.1,0.4) node[anchor=base,mydarkpurple ] {$1$};
\path (1.5,0.9) node[anchor=base,mydarkpurple] {$-1$};
 \draw [->,thick, mydarkorange ](0.5,-0.5)-- (-0.3,0.3); 
 \draw [->,thick, mydarkpurple ](0.5,-0.5)-- (1.4,-1.4); 
\path (-0.5, -0.4) node[anchor=base,mydarkorange] {$-\mathbf{\theta}$};
\path (0.9,-1.5) node[anchor=base ,mydarkpurple] {$\mathbf{\theta}$};
\path (4.9,1) node[anchor=base,mydarkpurple] {$a_{\theta} = 0.84$};
\path (4.9,0.5) node[anchor=base,mydarkorange] {$a_{-\theta} = 0.16$};
\fill(4.5,-0.5) circle (0.5mm);
\path (5.5,-0.5) node[anchor=base] {class $-1$};
\node[] at (4.5,-1.0) {*};
\path (5.4,-1.0) node[anchor=base] {class $1$};
\node[color = mydarkbrown] at (4.5,-1.4) {o};
\path (5.7,-1.5) node[anchor=base,mydarkbrown] {new input};
\end{tikzpicture}

\caption{Illustration of the decision boundary of a perceptron model for two-dimensional inputs. The parameter vector defines a hyperplane dividing input space into regions of class $-1$ and $1$. A change of sign of the parameters swaps the decision regions and leads to an accuracy of $a_{-\theta} = 1-a_{\theta}$, . }
\label{Fig:pointsymm}
\end{figure}
We will now turn to the question why the accuracies might be a good weighing scheme. Recall that there is a lot of evidence that ensembles of weak but accurate classifiers (meaning that $a_{\theta} > 0.5$ for all $\theta$) can lead to a strong classifier. The ensemble constructed in the last section however contains all sorts of models which did not undergo a selection procedure, and it may therefore contain a large --or even exponential-- share of models with low accuracy or random guessing. It turns out that for a large class of model families, the ensemble \textit{effectively} only contains accurate models. This to our knowledge is a new result also interesting for classical ensemble methods.\\

Assume the core machine learning model has the property to be point symmetric in the parameters $\theta$,
\[f(x; - \theta) = - f(x; \theta).\]
This is true for linear models and neural networks with an odd number of layers such as a simple perceptron or an architecture with $2$ hidden layers. Let us furthermore assume that the parameter space $\Theta$ is symmetric around zero, meaning that for each $\theta \in \Theta$ there is also $- \theta \in \Theta$. These pairs are denoted by $\theta^{+} \in \Theta^+, \theta^{-} \in \Theta^-$. With this notation one can write
\[f(x;\theta^{+}) = - f(x;\theta^{-}).  \]
From there it follows that $\int_{\theta \in \Theta} f(x; \theta) = 0$ and 
$a(\theta^+) = 1- a(\theta^{-})$. To get an intuition, consider a linear model imposing a linear decision boundary in the input space. The parameters define the vector orthogonal to the decision boundary (in addition to a bias term that defines where the boundary intersects with the y-axis which we ignore for the moment). A sign change of all parameters flips the vector around; the linear decision boundary remains at exactly the same position, meanwhile all decisions are turned around (see Figure \ref{Fig:pointsymm}).\\

\begin{figure}[t]
\centering
 \includegraphics[width=0.55\textwidth]{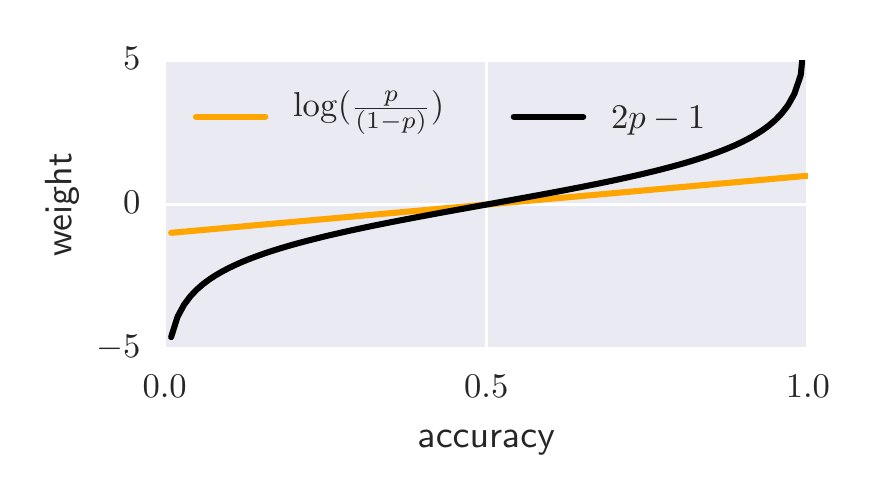}
 \caption{Comparison of the effective weight transformation described in the text and the optimal weight rule \cite{shapley84}. Both flip the prediction of a classifier that has an accuracy of less than $0.5$.}
\label{Fig:ens_shapley}
\end{figure}

For point symmetric models the expectation can be expressed as a sum over one half of the parameter space:
\begin{equation}
\frac{1}{E\chi} \sum\limits_{\theta} a_{\theta} f(\tilde{x},\theta)  = \frac{1}{E\chi} \sum\limits_{\theta^{+} } \left[ a(\theta^{+}) - \frac{1}{2} \right]  f(\tilde{x},\theta^{+}).
\label{Eq:expval}
\end{equation}
The result of this `effective transformation' is to shift the weights from the interval $[0,1]$ to $[-0.5,0.5]$, with profound consequences. The transformation is plotted in Figure \ref{Fig:ens_shapley}. One can see that accurate models get a positive weight, while non-accurate models get a negative weight and random guessers vanish from the sum. The negative weight consequently flips the decision $f(\x;\theta)$ of the `bad' models and turns them into accurate classifiers.  This is a linearisation of the rule mentioned in Eq. (\ref{Eq:quens_log}) as the optimal weight distribution for large ensembles (plotted in black for comparison).\\

With this in mind we can rewrite the expectation value as
\begin{equation}
\mathbb{E}[f(\tilde{x};\theta)] = \frac{1}{E} \sum\limits_{\theta | a(\theta) > 0.5 }  \tilde{a}(\theta)  f(\tilde{x},\theta),
\label{Eq:ensnewexp}
\end{equation}
with the new weights 
\begin{equation} \tilde{a}(\theta) = a(\theta) - \frac{1}{2},
\label{Eq:ensnewweights}
\end{equation}
 for parameters  $\theta^{+}$ as well as $\theta^{-}$.\\

If one cannot assume a point-symmetric parameter space but still wants to construct an ensemble of all accurate models (i.e., $a_{\theta} > 0.5$), an alternative to the above sketched routine $\mathcal{W}$ could be to construct an oracle that marks such models and then use amplitude amplification to create the desired distribution. A way to mark the models is to load every training vector into a predefined register and compute $\ket{f(\x^m; \theta)}$ as before, but perform a binary addition on a separate register that ``counts'' the number of correctly classified training inputs. The register would be a binary encoding of the count, and hence require $\lceil \log M \rceil$ qubits (as well as garbage qubits for the addition operation). If $\log M = \mu$ for an integer $\mu$, then the first qubit would be one if the number of correct classifications is larger than $M/2$ and zero otherwise. In other words, this qubit would flag the accurate models and can be used as an oracle for amplitude amplification. The optimal number of Grover iterations depends on the number of flagged models. In the best case, around $\frac{1}{4}$th of all models are accurate so that the optimal number of iterations is of the order of $\sqrt{E/\frac{1}{4}E} = 2$.  Of course, this number has to be estimated before performing the Grover iterations.\footnote{If one can analytically prove that  $\frac{1}{2}E$ of all possible models will be accurate as in the case of point-symmetric functions,  one can artificially extend the superposition to twice the size, prevent half of the subspace from being flagged and thereby achieve the optimal amplitude amplification scheme.} 

\section{Analytical investigation of the accuracy-weighted ensemble}\label{Sec6}

In order to explore the accuracy-weighted ensemble classifier further, we conduct some analytical and numerical investigations for the remainder of the article. It is convenient to assume that we know the probability distribution $p(x,y)$ from which the data is picked (that is either the `true' probability distribution with which data is generated, or the approximate distribution inferred by some data mining technique). Furthermore, we consider the continuous limit $\sum \rightarrow \int$. Each parameter $\theta$ defines decision regions in the input space, $\mathcal{R}^{\theta}_{-1}$ for class $-1$ and $\mathcal{R}^{\theta}_1$ for class $1$ (i.e. regions of inputs that are mapped to the respective classes). The accuracy can then be expressed as
\begin{equation}a(\theta) = \frac{1}{2}\int\limits_{\mathcal{R}^{\theta}_{\text{-}1}} p(x, y \seq \text{-}1) \;dx + \frac{1}{2}\int\limits_{\mathcal{R}^{\theta}_1} p(x, y \seq 1) \;dx.\label{Eq:ensanalacc}\end{equation}
In words, this expression measures how much of the density of a class falls into the decision region proposed by the model for that class. Good parameters will propose decision regions that contain the high-density areas of a class distribution. The factor of $1/2$ is necessary to ensure that the accuracy is always in the interval $[0,1]$ since the two probability densities are each normalised to $1$.\\

The probability distributions we consider here will be of the form $p(x, y=\pm1) = g(x; \mu_{\pm}, \sigma_{\pm}) = g_{\pm}(x) $. They are normalised, $\int_{-\infty}^{\infty} g_{\pm}(x) dx =1$, and vanish asymptotically, $g_{\pm}(x) \rightarrow 0$ for $x \rightarrow -\infty$ and $x \rightarrow +\infty$. The hyperparameters $\mu_{-}, \sigma_{-}$ and $\mu_{+}, \sigma_{+}$ define the mean or `position' and the variance or `width' of the distribution for the two classes $-1$ and $1$. Prominent examples are Gaussians or box functions. Let $G(x; \mu_{\pm}, \sigma_{\pm}) = G_{\pm}(x)$ be the integral of $g(x; \mu_{\pm}, \sigma_{\pm})$, which fulfils  $ G_{\pm}(x) \rightarrow 0$ for $x \rightarrow -\infty$ and $ G_{\pm}(x) \rightarrow 1$ for $x \rightarrow +\infty$. Two expressions following from this property which will become important later are
\[\int\limits^{a}_{-\infty} g_{\pm}(x) dx =  G_{\pm}(a) - G_{\pm}(-\infty) =G_{\pm}(a),\]
and
\[\int\limits_{a}^{\infty} g_{\pm}(x) dx = G_{\pm}(\infty) - G_{\pm}(a) =1 - G_{\pm}(a).\]

\begin{figure}[t]
\centering
\begin{tikzpicture}{scale = 0.8}
\draw [->, thick](-2,0)-- (0,0); 
\draw [->, thick](1,0)-- (3,0); 

\path (-0.1,-0.3) node[anchor=base] {$x$};
\path (3,-0.3) node[anchor=base] {$x$};

\draw[dotted, thick] (-1,-0.2) -- (-1,2.1);
 \draw [-> ](-1,1)-- (-1.5, 1); 
 \draw [-> ](-1,1)-- (-0.5, 1); 
\path (-1.8,1.2) node[anchor=base ] {class $-1$};
\path (-0.2,1.2) node[anchor=base ] {class $1$};
\path (-1,-0.7) node[anchor=base] {$((w_0)_1, o = 1)$};

\draw[dotted, thick] (2,-0.2) -- (2,2.1);
 \draw [-> ](2,1)-- (2.5, 1); 
 \draw [-> ](2,1)-- (1.5, 1); 
\path (2.8,0.5) node[anchor=base ] {class $-1$};
\path (1.2,0.5) node[anchor=base ] {class $1$};
\path (2,-0.7) node[anchor=base] {$((w_0)_2, o = -1)$};
\end{tikzpicture}
\caption[Simplified $1$-d perceptron]{Very simple model of a $1$-dimensional classifier analysed in the text: The parameter $w_0$ denotes the position of the decision boundary while parameter $o$ determines its orientation.}
\label{Fig:quens_db_principle}
\end{figure}
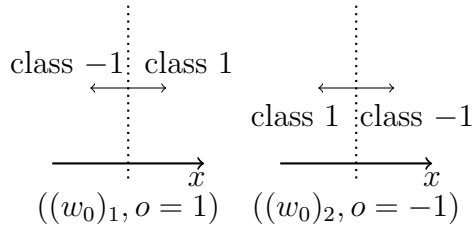

We consider a minimal toy example for a classifier, namely a perceptron model on a one-dimensional input space, $f(x; w,w_0) = \mathrm{sgn}(wx + w_0)$ with $x,w,w_0 \in \mathbb{R}$. While one parameter would be sufficient to mark the position of the point-like `decision boundary', a second one is required to define its orientation. One can simplify the model even further by letting the bias $w_0$ define the position of the decision boundary and introducing a binary `orientation' parameter $o \in \{-1,1\}$ (as illustrated in Figure \ref{Fig:quens_db_principle}),
\[ f(x; o ,w_0) = \mathrm{sgn}( o (x - w_0) ). \]
For this simplified perceptron model the decision regions are given by
\[\mathcal{R}_{-1} = [-\infty, w_0], \mathcal{R}_{1} = [w_0,\infty] ,\]
for the orientation $o=1$  and 
\[\mathcal{R}_{-1} = [w_0,\infty] , \mathcal{R}_{1} = [-\infty, w_0] ,\]
for $o=-1$. 
Our goal is to compute the expectation value 
\begin{equation}
\mathbb{E}[f(\tilde{x};w_0,o)] \propto \int d\theta\; a(\theta) f(\tilde{x}; o,w_0) ,
\label{Eq:expval}
\end{equation}
of which the sign function evaluates the desired prediction $\tilde{y}$. \\

Inserting the definitions from above, as well as some calculus using the properties of $p(x)$ brings this expression to
\begin{equation*}   \int\limits_{-\infty}^{\infty} dw_0\; ( G_-(w_0) -  G_+(w_0))  \;\mathrm{sgn}(\tilde{x}-w_0),
\end{equation*}
and evaluating the sign function for the two cases $\tilde{x} > w_0, \tilde{x} < w_0 $ leads to  
\begin{equation}   \int\limits_{-\infty}^{\tilde{x}} dw_0  \Big\{ G_+(w_0) -G_-(w_0)  \Big\} - \int\limits_{\tilde{x}}^{\infty} dw_0  \Big\{ G_+(w_0) -  G_-(w_0) \Big\} .
\label{Eq:expr}
\end{equation}

\begin{figure}[t]
\centering
       \includegraphics[width=0.5\textwidth]{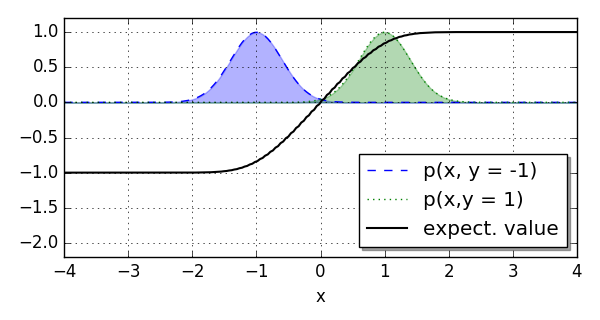}
 \caption{The black line plots the expectation value of the accuracy-weighed ensemble decision discussed in the text for the two class densities and for different new inputs $x$. A positive expectation value yields the prediction $1$ while a negative expectation value predicts class $-1$. At $\mathbb{E}[f(x;w_0,o)]  = 0$ lies the decision boundary. The plot shows that the model predicts the decision boundary where we would expect it, namely exactly between the two distributions. }
\label{Fig:quens_expvalvsdistr}
\end{figure}

\begin{figure}[t]
\centering
       \includegraphics[width=0.35\textwidth]{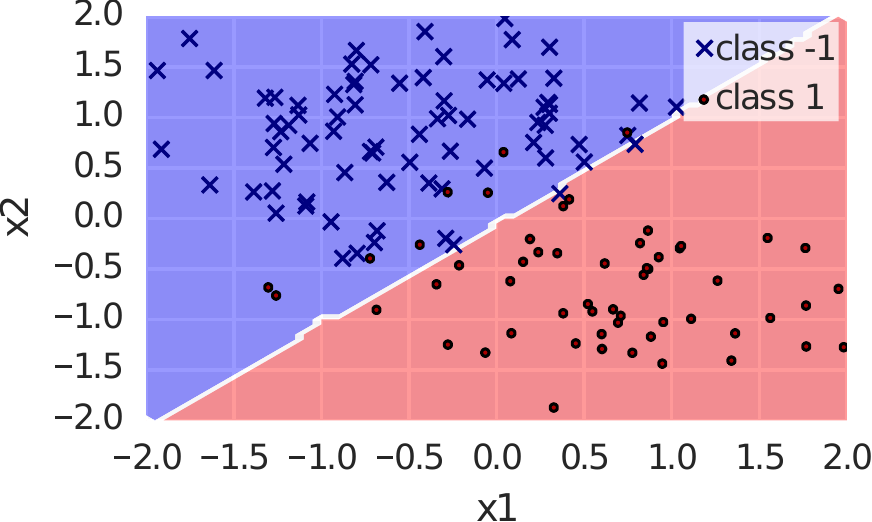}
 \caption[Simulations for an ensemble of perceptron classifiers]{Perceptron classifier with $2$-dimensional inputs and a bias. The ensemble consist of  $8000$ models, each of the three parameters taking $20$ equally distributed values from $[-1,1]$. The resolution to compute the decision regions was $\Delta x_1 = \Delta x_2 = 0.05$. The dataset was generated with python scikit-learn's blob function and both classes have the same variance parameter. One can see that in two dimensions, the decision boundary still lies in between the two means of $\mu_{-1} = [-1,1] $ and $\mu_{1} = [1,-1]$.}
\label{Fig:quens_2dperc}
\end{figure}
To analyse this expression further, consider the two class densities to be Gaussian probability distributions
\[p(x, y = \pm 1) = \frac{1}{\sigma_{\pm} \sqrt{2\pi}} \exp^{-\frac{(x-\mu_{\pm})^2}{2\sigma_{\pm}^2}}.\]
The indefinite integral over the Gaussian distributions is given by
\[G_{\pm}(x) = \frac{1}{2} [1 + \mathrm{erf}\frac{ x-\mu_{\pm}}{\sqrt{2}\sigma_{\pm}}]. \]
Inserting this into Eq (\ref{Eq:expr}) we get an expectation value of
\begin{equation}
\int\limits_{-\infty}^{\tilde{x}} dw_0  \Big\{\mathrm{erf}\frac{ w_0-\mu_+}{\sqrt{2}\sigma_+} -  \mathrm{erf}\frac{ w_0-\mu_-}{\sqrt{2}\sigma_-}  \Big\} -\int\limits_{\tilde{x}}^{\infty} dw_0  \Big\{  \mathrm{erf}\frac{ w_0-\mu_+}{\sqrt{2}\sigma_+}- \mathrm{erf}\frac{ w_0-\mu_-}{\sqrt{2}\sigma_-}\Big\}.  
\label{Eq:quens_expval}
\end{equation}
Figure \ref{Fig:quens_expvalvsdistr} plots the expectation value for different inputs $\tilde{x}$ for the case of two Gaussians with $ \sigma_-= \sigma_+ = 0.5$, and $\mu_- = -1 , \mu_+ = 1 $. The decision boundary is at the point where the expectation value changes from a negative to a positive value, $\mathbb{E}[f(\hat{x};w_0,o)] = 0 $. One can see that for this simple case, the decision boundary will be in between the two means, which we would naturally expect. This is an important finding, since it implies that the accuracy-weighted ensemble classifier works - arguably only for a very simple model and dataset. A quick calculation shows that we can always find the decision boundary midway between the two means if $ \sigma_-= \sigma_+$. In this case the integrals in Eq. (\ref{Eq:quens_expval}) evaluate to 
\[ \int\limits_{-\infty}^{\tilde{x}} dw_0 \; \mathrm{erf}\frac{ w_0-\mu_{\pm}}{\sqrt{2}\sigma_{\pm}} =  \gamma_{\pm}(\tilde{x}) - \lim_{R\rightarrow -\infty}  \gamma_{\pm}(R) ,\]
and
\[ \int\limits_{\tilde{x}}^{\infty}dw_0 \; \mathrm{erf}\frac{ w_0-\mu_{\pm}}{\sqrt{2}\sigma_{\pm}} = \lim_{R\rightarrow \infty}  \gamma_{\pm}(R)  -  \gamma_{\pm}(\tilde{x}) , \] 
with the integral over the error function
\[\gamma_{\pm}(x)  = (x - \mu_{\pm})\; \mathrm{erf}\left(\frac{x-\mu_{\pm}}{\sqrt{2}\sigma_{\pm}}\right) +  \sqrt{\frac{2}{\pi}} \sigma_{\pm} \mathrm{e}^{- (\frac{ x-\mu_{\pm}}{\sqrt{2}\sigma_{\pm}} )^2}    . \]
Assuming that the mean and variance of the two class distributions are of reasonable (i.e. finite) value, the error function evaluates to $0$ or $1$ before the limit process becomes important, and one can therefore write
\begin{eqnarray}
\lim_{R\rightarrow -\infty}  \gamma_{\pm}(R) &=  \lim_{R\rightarrow -\infty}   \sqrt{\frac{2}{\pi}} \sigma_{\pm} \mathrm{e}^{- (\frac{R-\mu_{\pm}}{\sqrt{2}\sigma_{\pm}} )^2},\\
\lim_{R\rightarrow \infty}  \gamma_{\pm}(R)   &=  \lim_{R\rightarrow \infty}   R \; +  \lim_{R\rightarrow \infty}  \sqrt{\frac{2}{\pi}} \sigma_{\pm} \mathrm{e}^{- (\frac{R-\mu_{\pm}}{\sqrt{2}\sigma_{\pm}} )^2} - \mu.
\end{eqnarray}  
The expectation value for the case of equal variances therefore becomes
\begin{equation}
\mathbb{E}[f(\tilde{x};w_0,o)]  =   2\gamma_{-}(\tilde{x})    -2 \gamma_{+}(\tilde{x}) .  
\label{Eq:Eq:toyexpval} 
\end{equation}
Setting $\tilde{x} = \hat{\mu} =\mu_- + 0.5(\mu_+ + \mu_-)$ turns the expectation value to zero; the point $\hat{\mu} $ between the two variances is shown to be the decision boundary. Simulations confirm that this is also true for other distributions, such as a square, exponential or Lorentz distribution, as well as for two-dimensional data (see Figure \ref{Fig:quens_2dperc}).\\

\begin{figure*}[t]
\centering
Example 1\\
       \includegraphics[width=0.7\textwidth]{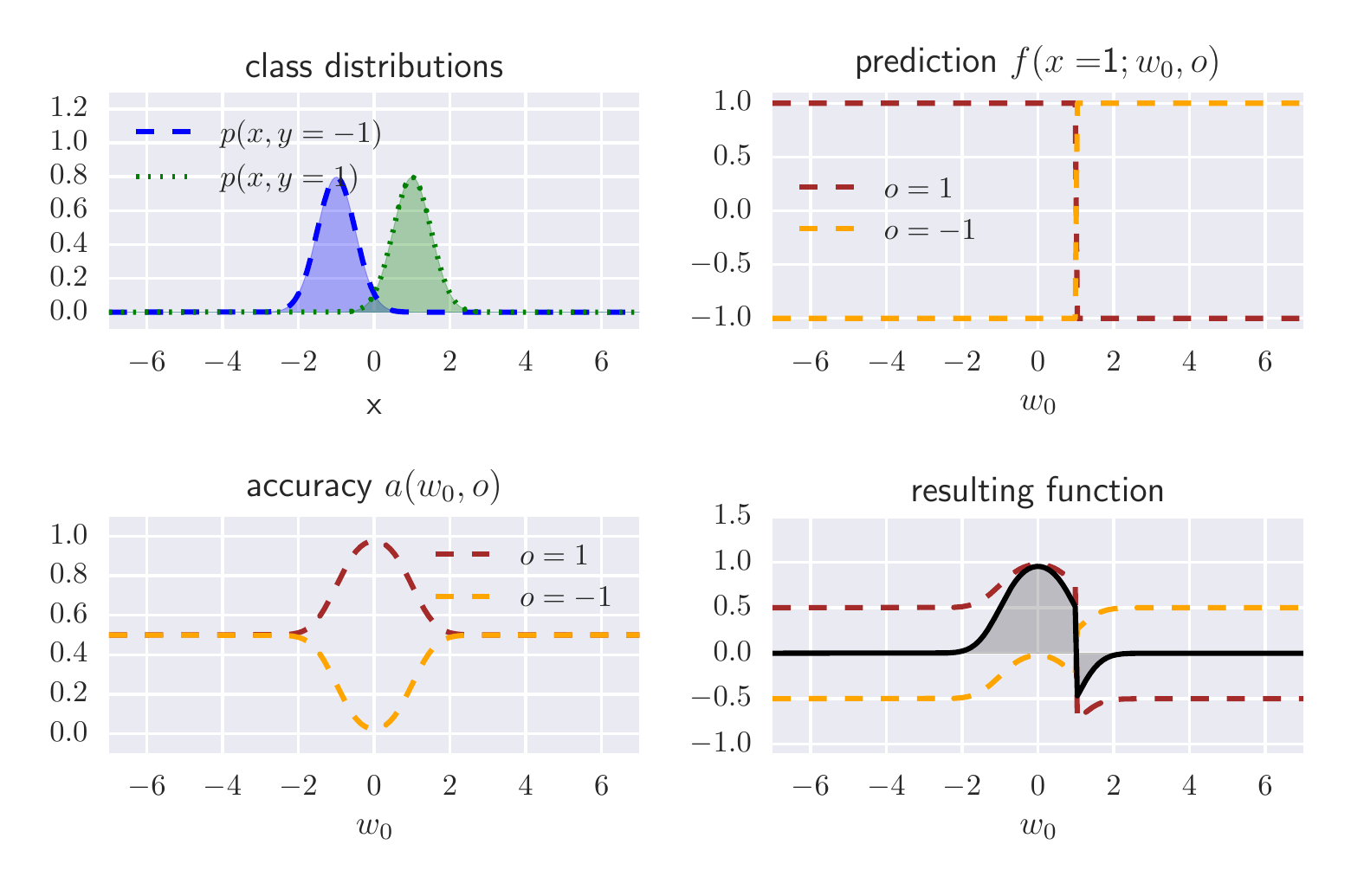}   \\    
Example 2\\
\includegraphics[width=0.7\textwidth]{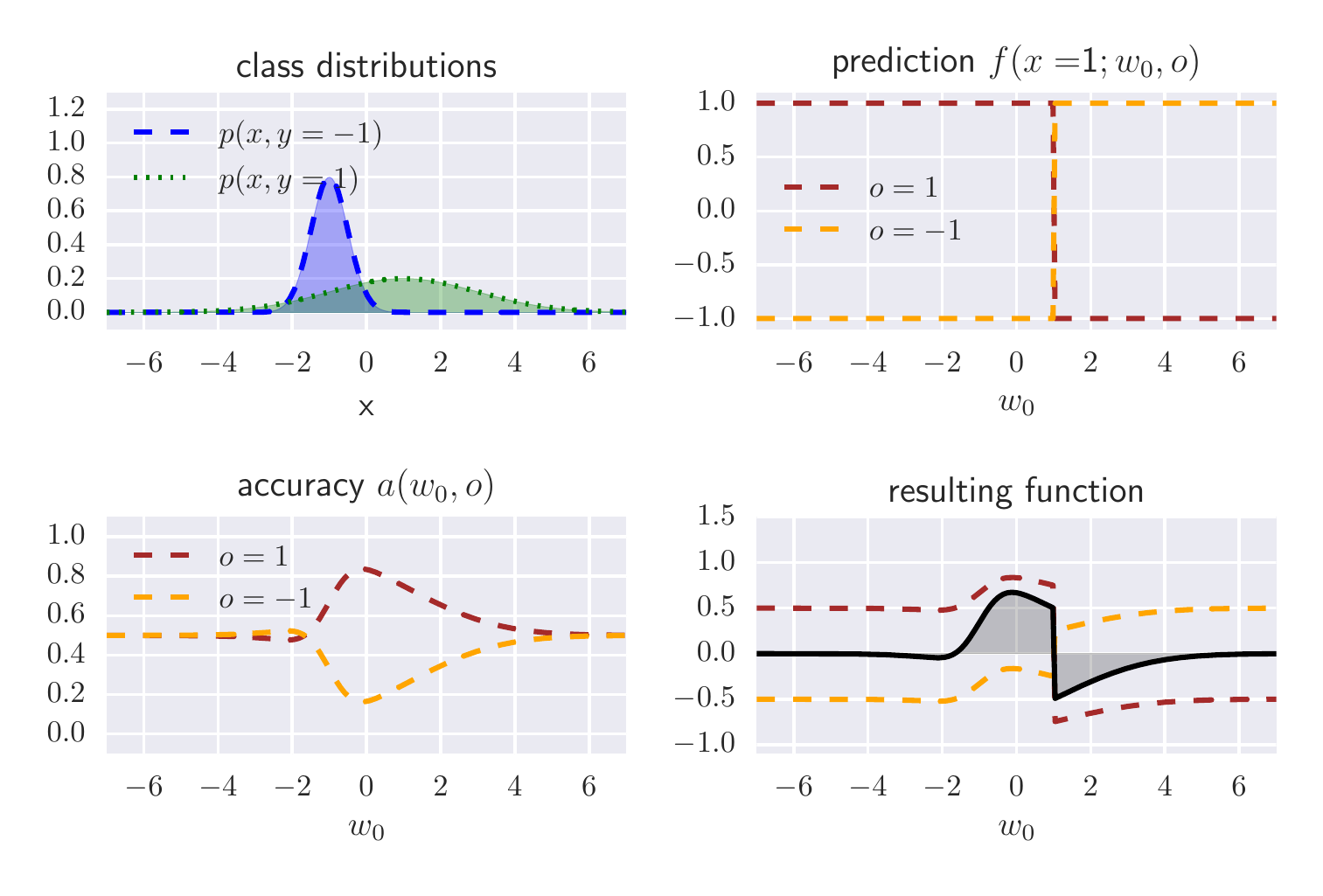}
 \caption[Detailed analysis of the ensemble classifier]{Detailed analysis of the classification of $\tilde{x} = 1$ for two examples of normal data distributions. The upper Example 1 shows $p(x, y = -1) = \mathcal{N}(-1, 0.5)$ and $p(x, y = 1) = \mathcal{N}(1, 0.5)$ while the lower Example 2 shows $p(x, y = -1) = \mathcal{N}(-1, 0.5)$ and $p(x, y = 1) = \mathcal{N}(1, 2)$ (plotted each in the top left figure of the four). The top right figure in each block shows the classification of a given new input $\tilde{x}$ for varying parameters $b$, plotted for $o=1$ and $o = -1$ respectively. The bottom left shows the accuracy or classification performance $ a(w_0, o = 1) $ and  $a(w_0, o = -1)$ on the data distribution. The bottom right plots the product of the previous two for $o= 1$ and $o = -1$, as well as the resulting function under the integral. The prediction outcome is the integral over the black curve, or the total of the gray shaded areas. One can see that for models with different variances, the accuracies loose their symmetry and the decision boundary will therefore not lie in the middle between the two means.  }
\label{Fig:quens_minclassdetails}
\end{figure*}

The simplicity of the core model allows us to have a look into the structure of the expectation value. Figure \ref{Fig:quens_minclassdetails} shows the components of the integrand in Eq. (\ref{Eq:expval}) for the expectation value, namely the accuracy, the core model function as well as their product. Example 1 shows the same variances $\sigma_+ = \sigma_-$, while Example 2 plots different variances. The plots show that for equal variances, the accuracy is a symmetric function centred between the two means, while for different variances, the function is highly asymmetric. In case the two distributions are sufficiently close to each other, this has a sensitive impact on the position of the decision boundary, which will be shifted towards the flatter distribution. This might be a desired behaviour in some contexts, but is first and foremost an interesting result of the analytical investigation. \\

As a summary, analysing a very simple classifier in relation to one-dimensional Gaussian data distributions gives evidence that the weighted average ensemble classifier can lead to meaningful predictions for separated data, but there seems to be a sensitive dependency on the shape of the distributions. Further investigations would have to show how the accuracy-weighted ensemble classifier behaves with more complex base classifiers and/or realistic datasets. Low resolution simulations with one-dimensional inputs confirm that nonlinear decision boundaries can in fact be learnt. However, computing the exact expectation value is a huge computational effort. For example, the next more complex neural network model requires two hidden layers to be point symmetric, and with one bias neuron and for two-dimensional inputs the architecture has already seven or more weights. If each weight is encoded in three qubits only (including one qubit for the sign), we get an ensemble of $2^{21}$ members whose collective decision needs to be computed for an entire input space in order to determine the decision boundaries.  Sampling methods could help to obtain approximations to this result, and would open these methods to classical applications as well.

\section{Conclusion}\label{Sec7}
This article proposed a framework to construct quantum ensembles of quantum classifiers which use parallelism in order to evaluate the predictions of exponentially large ensembles. The proposal leaves a lot of space for further research. First, as mentioned above, the quantum ensemble has interesting extensions to classical methods when considering approximations to compute the weighted sum over all its members' predictions. Recent results on the quantum supremacy of Boson sampling show that the distributions of  some state preparation routines (and therefore some quantum ensembles), cannot be computed efficiently on classical computers. Are there meaningful quantum ensembles for which a speedup can be proven? Are the rules of combining weak learners to a strong learner different in the quantum regime? Which types of ensembles can generically prepared by quantum devices?  Second, an important issue that we did not touch upon is overfitting. In AdaBoost, regularisation is equivalent to early stopping \cite{schapire13}, while Bayesian learning has inbuilt regularisation mechanisms. How does the accuracy-based ensemble relate to these cases? Is it likely to overfit when considering more flexible models? A third question is whether there are other (coherent) ways of defining a quantum ensemble for quantum machine learning algorithms that do not have the format of a quantum classifier as defined above. Can mixed states be used in this context? In summary, this article provided a first step to think about ensemble methods in quantum machine learning and is an example for the mutual enrichment that classical and quantum machine learning can provide for each other.

\section*{Acknowledgments}
This work was supported by the South African Research Chair Initiative of the Department of Science and Technology and the National Research Foundation. We thank Ludmila Kuncheva and Thomas Dietterich for their helpful comments.\bigbreak


\end{document}